\documentclass[%
 aip,
 apl,%
 amsmath,amssymb,
 reprint,%
]{revtex4-1}

\usepackage{graphicx}
\usepackage{dcolumn}
\usepackage{bm}
\usepackage{times}
\usepackage[belowskip=-15pt,aboveskip=0pt]{caption}
\usepackage{color}
\usepackage{epsfig}

\definecolor{darkgreen}{rgb}{0,0.5,0}
\definecolor{purple}{rgb}{1,0,1}

\makeatletter
\def\ScaleIfNeeded{%
\ifdim\Gin@nat@width>\linewidth
\linewidth
\else
\Gin@nat@width
\fi
}
\RequirePackage[normalem]{ulem} 
\RequirePackage{color}\definecolor{RED}{rgb}{1,0,0}\definecolor{BLUE}{rgb}{0,0,1} 

\begin{document}

\preprint{AIP/123-QED}

\title{Improved Efficiency of Plasmonic Tin Sulfide Solar Cells}

\author{Priyal Jain}
\affiliation{ 
Department of Electronic Science, University of Delhi, 
South Campus, Delhi 110 021, India.
}%
\author{Arun P}%
\email{arunp92@physics.du.ac.in}
\affiliation{ 
Material Science Research Lab, S.G.T.B. Khalsa College,University of Delhi, Delhi, 110 007, India.
}%

\date{\today}

\begin{abstract}
Solar cells with the structure ITO-PEDOT:PSS-Ag:SnS-Al were fabricated with the active layer of tin sulphide with silver nano-particles (Ag:SnS) grown by thermal co-evaporation. To understand the influence of the silver nanoparticles on the energy conversion process, various cells with varying active layer thicknesses were compared. Results showed that the Ag nanoparticles act as scattering centers, resulting in longer optical path lengths for incident light. This in turn results in more charge carriers being generated and thus enhances the efficiency of the structure as compared to the pristine ITO-PEDOT:PSS-SnS-Al structure. The plasmonic solar cells of SnS showed an improvement of more than 40\%. The results are encouraging and suggests more concerted effort needs to be made on SnS.
\end{abstract}

\pacs{71.20.Ps, 84.60.Jt, 75.50.Tt}
\keywords{Thin Films, Nanoparticles, Plasmonic Solar Cells}
\maketitle

Present commercial solar cells are silicon wafer based with standard
efficiency around 24.7\%.~\cite{silicon} However, using the single crystal wafers cut to micro-meter thicknesses, increases the cost and makes 
utilization of solar energy economically unattractive. Researchers hence 
have started looking into organic~\cite{organic1,organic2} and inorganic~\cite{cds,arbele} materials in thin film state either in amorphous or 
polycrystalline state.~\cite{amp,poly} The hope is to increase efficiency 
at a lower cost, effectively decreasing the energy conversion cost. The
inorganic solar cells have not matched the efficiency of Silicon wafer solar
cells as yet, with the best efficiency reported at 16.5\% for Cadmium
Telluride (CdTe)~\cite{cdte} film solar cells. However, with concerns on the 
toxicity of tellurium used, search has moved to other inorganic materials 
such as Tin Sulphide (SnS)~\cite{reddy,david} and lead sulphide (PbS).~\cite{pbs}

The basic idea of solar cells are to absorb photons and generate charge
carriers within the junction which should readily separate and reach their
respective electrodes without recombination. To improve efficiency one requires a good absorbing material for the photo-active layer. This can be achieved either by using an active layer of good absorbance or increase its thickness.~\cite{david} However, the thickness of the layer is limited by the carrier diffusion length, i.e. the thickness of the active layer should always be less than the carrier diffusion length.~\cite{atwater} If the 
thickness of the active layer is larger than the carrier diffusion length, 
the carriers would recombine before being collected at the electrodes. This
would reduce the current in the external circuitry and hence diminish the 
performance of the cell. Also, increasing the dimension of the absorber layer 
would raise the cost of the device. Selection of the second layer with
opposite charge carriers (n or p type) to that of the semiconducting nature
of the photo-active layer (p or n type) also plays an important role. The selection is made in view of matching the conduction and valence energy
levels of the two layers such that the separation of the generated charge carriers is encouraged.

To increase the absorbance of the active layer without increasing the
thickness, researchers are now investigating effects of embedding metal nanoparticles within the active layer. When light is incident on the metal/dielectric or a metal/semiconductor interface, it gives rise to collective oscillation of electrons known as ``Surface Plasmon Resonance (SPR)".~\cite{spr} At this particular resonance frequency, the nanoparticles either effectively absorbs light or scatters it.~\cite{spr1} The amount of light scattered or absorbed depends on the size of the metal nanoparticles.~\cite{spr1,spr2} Solar cells that use metal nano-clusters are called plasmonic solar cells. Derkacs et al.~\cite{derkacs} fabricated thin film amorphous Silicon solar cells containing
Au nanoparticles and obtained an 8.3\% increase in the conversion efficiency, while Schaadt et al.~\cite{schaadt} reported an enhancement of 80\% when Au nanoparticles were incorporated in crystalline Silicon solar cells. 
Recently, Pillai et al.~\cite{pillai} reported a 30\% enhancement in absorption in crystalline Silicon solar cells including Ag nanoparticles. 
Similar, studies were also carried out on other semiconductor solar cells 
as well. Konda et al.~\cite{konda} reported a significant enhancement in the photo-current for n-CdSe heterojunction solar cells containing Au nanoparticles. Stenzel et al.~\cite{stenzel} reported an improvement 
in photo-current by a factor of 2.7 when Au or Cu nanoparticles were embedded into ITO-Copper Phthalocyanine-In structures. Similar studies have been made in organic solar cells too,~\cite{org1,org2} however till date, this plasmonic enhancement approach has not seen in SnS thin films. Thus, it would be interesting to see if the formation of LSPR improves the efficiencies of SnS solar cells. 

SnS based solar cells today have very low conversion efficiency (at best 2\%~\cite{prasert}). Various research groups hence have focused on the factors that can improve the efficiency of SnS solar cells~\cite{david},\cite{ghosh,noguchi}. While majority of research try to address issue of selecting an appropriate companion `p' or `n'-layer,~\cite{ristov,alex} few groups have focused on improving the absorption of the photo-active SnS layer,~\cite{david} by varying the thickness,~\cite{david} or as in the stray example by embedding metal clusters within the film thickness~\cite{ristov} (However, it must be noted that Ristov et al did not talk of SPR in their work).

\begin{figure}[h!!]
\epsfig{file=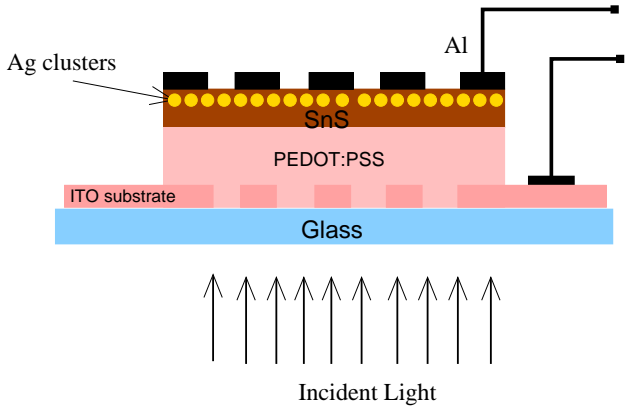, width=3.25in}
\vskip 0.3cm
\caption{Figure shows the structure of our solar cell with layers of 
ITO/PEDOT:PSS/SnS:Ag/Al. The circles in the figure are the Ag nanoparticles 
that reside at the top surface of SnS films.}
\vskip 0.3cm
\label{experiment1}
\end{figure}

Tan et al~\cite{tan} in 2010 reported a solar cell structure based on ${\rm n-SnS_2}$ quantum dots with PEDOT:PSS on ITO substrate. Both the PEDOT:PSS and ${\rm n-SnS_2}$ layers were fabricated by spin-coat technique. We were able to fabricate this solar cells by thermally evaporating SnS on PEDOT:PSS-ITO acting as the p-type layer. We shall henceforth refer to this structure (ITO-PEDOT:PSS-SnS-Al) as the pristine solar cell. By modeling the solar cell as a single diode circuit, the cell parameters such as series resistance (${\rm R_s}$), shunt resistance (${\rm R_p}$), open circuit voltage (${\rm V_{oc}}$), Fill Factor (FF), ideality factor (n) and efficiency (${\rm \eta}$) were determined. In terms of these parameters, we were able to identically reproduce the solar cell reported by Tan et al~\cite{tan} with a maximum
efficiency of 0.21\% for a cell whose SnS film thickness was 
800~nm and grain size around 17.7~nm.~\cite{solarpaper}

To increase the efficiency of these solar cells, we decided to fabricate
plasmonic solar cells of SnS using 20~nm silver nano-particles.
Fig~1 shows modifications made in our pristine solar cell device which
incorporates oblate metal nano-particles of Ag residing at the surface of
SnS film away from SnS/PEDOT:PSS junction.
This article reports the performance of our plasmonic solar cells and
compares its parameters with those of pristine SnS solar cells.

\section{Experimental Details}

Solar cell structures of n-SnS:Ag/PEDOT:PSS were fabricated on etched Indium 
Tin Oxide (ITO) substrates of low resistivity  (${\rm 10-15~\Omega}$/sq). A
200~nm layer of aqueous solar grade PEDOT:PSS (1.3~\%) was spin coated on 
the substrates. Followed by this, composite thin films of Tin sulfide (SnS) 
and silver (Ag) were grown on the PEDOT:PSS layer. The films were grown at
room temperature by thermal evaporation technique using Hind Hi-Vac (12A4D)
coating unit at vacuum better than ${\rm 4 \times 10^{-5}}$~Torr. Before
evaporation, pellets were made by mixing SnS powder and Ag nano-powder. The 
SnS powder (99.99~\% pure) was provided by Himedia (Mumbai) and the Ag 
nano-powder was provided by Nanoshel (USA). For mixing, the (mass) ratio of 
SnS:Ag taken was in a proportion of 2:1. The thickness of the composite
films were measured using Veeco’s Dektak Surface profiler (150) on films 
grown simultaneously on glass substrates. Finally, Aluminum electrodes 
were deposited by thermal evaporation using standard masks. The thicknesses of 
the SnS:Ag composite films were varied while that of the spin coated 
PEDOT:PSS was maintained fixed.

The structural characterizations of the SnS films were done using a Bruker D8 
diffractometer at an operating voltage of 40~KV in the ${\rm \theta-2\theta}$ mode with Cu target giving X-Ray of ${\rm \lambda=1.5416~\AA}$. The 
current-voltage (J-V) measurements were done with a computer monitored 
Keithley 2400 source meter unit. A solar simulator of ${\rm 100~mW/cm^2}$, 
Air Mass (AM) 1.5 spectrum was used as illumination source. The measurements 
were made with light incident from the ITO side. 

\section {Results and discussion}
In our previous study \cite{jap} we showed that Ag nano-clusters
embedded in SnS acts as center for light scattering thus increasing the
optical path length of light within the active layer. As stated, this is 
expected to increase the efficiency of the solar-cell. In the following 
passages we investigate this and compare the photo-voltaic parameters of 
our plasmonic-device with pristine SnS solar cells. 

\begin{figure}[h!!]
\epsfig{file=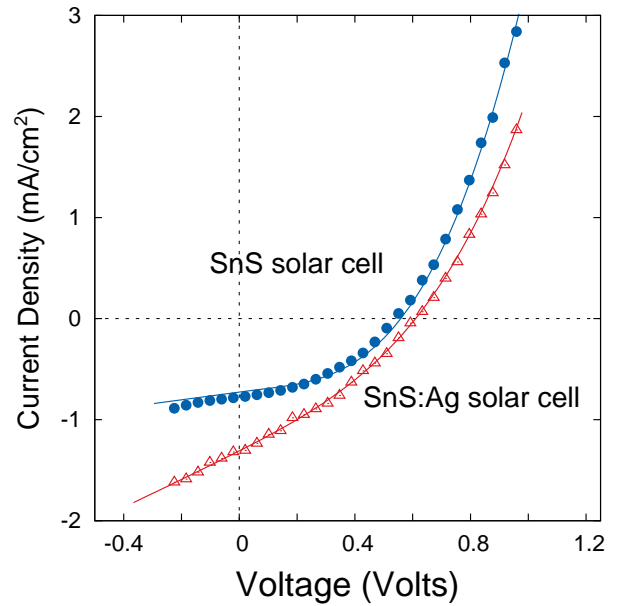, width=3.25in, angle=-90}
\vskip 0.3cm
\caption{A comparison of J-V characteristics of SnS and SnS:Ag plasmonic 
solar cells for active layers of 800~nm.}
\vskip 0.3cm
\label{jvnew2}
\end{figure}

Fig~2 compares the J-V characteristics of SnS solar cell
with that of a plasmonic SnS:Ag solar cell. The curves selected are of cells
with 800~nm thick photo-active layer and are representative of all measurements made. An increase in the efficiency of the plasmonic SnS:Ag cell can be judged by the increase in area enclosed by the J-V curve in the forth quadrant. In our study on pristine cells,~\cite{solarpaper} we had discussed the properties as a function of SnS grain size. However, the analysis of our previous work~\cite{jap} suggests that Ag nanoparticle's efficiency to scatter light into the dielectric SnS background depends on the Ag nanoparticles's grain size, it would be necessary to compare conversion efficiencies as a function of a parameter on which both Ag and SnS grain size would depend on. Fortunately, both SnS grain size~\cite{jos} and Ag nanoparticle's grain size~(fig~3) show a linear trend with the active layer's thickness (within region of experimental interest).

\begin{figure}[h!!]
\epsfig{file=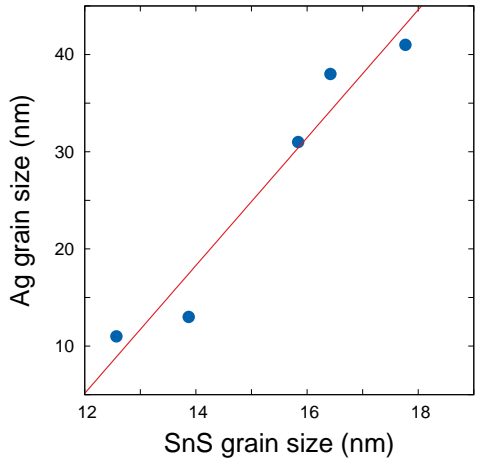, width=2.5in, angle=-90}
\vskip 0.3cm
\caption{The variation in average Ag nanoparticles size with backgrounds SnS grain size.}
\vskip 0.3cm
\label{grainrel}
\end{figure}

Hence, in fig~4, we plot the efficiency of SnS thin film solar cells and SnS:Ag plasmonic Solar cells as a function of the active layer's thickness. The conversion efficiency increases with increasing film thickness. In fact, for an active layer of thickness ~800nm, 
the conversion efficiency of the device is enhanced by 42$\rm\%$ when the Ag 
nanoparticles were introduced in SnS films. The increase (rate) in efficiency with film thickness can not be explained merely by the insignificant variation in absorbance with film thickness reported.~\cite{tsf,jos} 

\begin{figure}[h!!]
\epsfig{file=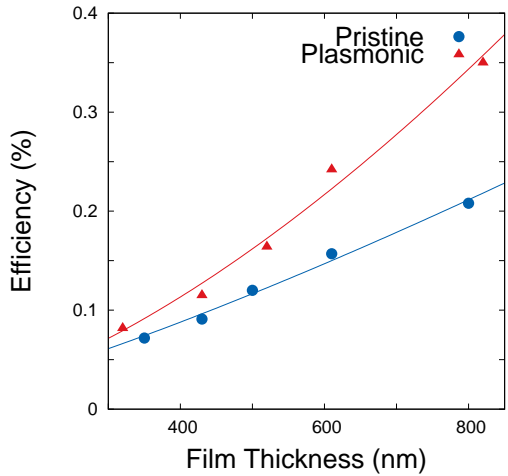, width=2.5in, angle=-90}
\vskip 0.3cm
\caption{A comparison of efficiencies of pristine SnS and SnS:Ag plasmonic 
solar cells for various active layer's thickness.}
\label{eff}
\vskip 0.3cm
\end{figure}

The increase in efficiency is hence related to the scattering from Ag nanoparticles. Larger metal nanoparticles results in more scattering. This increases the optical path of incident light within the cell leading to more charge carriers being released. A relook at fig~2 shows that the increase in efficiency of Ag:SnS solar cell is due to the increasing short-circuit current density, ${\rm J_{sc}}$ (point at which J-V curve cuts the `Y'-axis) with marginal or no increase in open circuit voltage, ${\rm V_{oc}}$ (point at which J-V curve cuts the `X'-axis). Fig~5 shows a plot between Ag nanoparticle's grain size and ${\rm J_{sc}}$. The trend confirms that more charge carriers are generated with increasing Ag nanoparticle's size.  A maximum enhancement of 
27$\rm\%$ in the photo-current was measured in SnS:Ag solar cells as compared to the pristine cells. 

\begin{figure}[h!!]
\epsfig{file=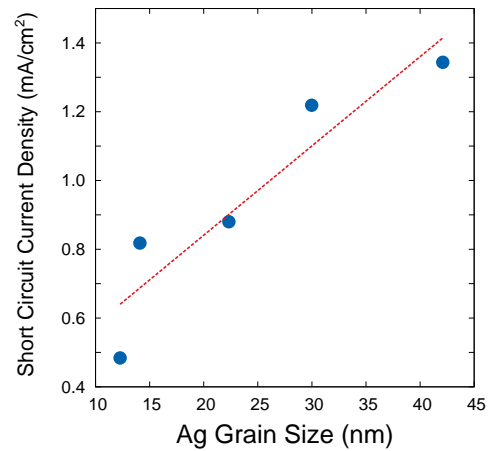, width=2.5in, angle=-90}
\caption{Variation of the solar cell's short circuit current, ${\rm I_{sc}}$, with Ag nanoparticle's average grain size.}
\label{new}
\end{figure}

\section{Conclusion}
Plasmonic solar cells of SnS were fabricated by co-evaporation of silver metal during the fabrication of ITO-PEDOT:PSS-Ag:SnS-Al structures. A substantial increase in the conversion efficiency of the device was observed as compared to the pristine (non-plasmonic) ITO-PEDOT:PSS-SnS-Al structures. The increase in efficiency is shown to be due to increasing photo-current generated due to increase in light scattering within the cell due to the Ag nanoparticles. The results are promising considering that we now can manipulate the inorganic, non-toxic SnS solar devices for higher efficiency.

\end{document}